\global\def\draftcontrol{0}

   \def\versionno{ cw sppression -- draft   }

\catcode`\@=11

\expandafter\ifx\csname draftcontrol\endcsname\relax\global\def\draftcontrol{0}
\fi

{\count255=\time\divide\count255 by 60
\xdef\hourmin{\number\count255}
\multiply\count255 by-60\advance\count255 by\time
\xdef\hourmin{\hourmin:\ifnum\count255<10 0\fi\the\count255}}
\def\draftdate{\number\month/\number\day/\number\year\ \ \ \hourmin }

\newcommand\makepapertitle{\par
  \begingroup
    \renewcommand\thefootnote{\@fnsymbol\c@footnote}%
    \def\@makefnmark{\rlap{\@textsuperscript{\normalfont\@thefnmark}}}%
    \long\def\@makefntext##1{\parindent 1em\noindent
            \hb@xt@1.8em{%
                \hss\@textsuperscript{\normalfont\@thefnmark}}##1}%
     \newpage
     \global\@topnum\z@   
     \@makepapertitle
     \thispagestyle{empty}\@thanks
  \endgroup
  \setcounter{footnote}{0}%
  \global\let\thanks\relax
  \global\let\makepapertitle\relax
  \global\let\@makepapertitle\relax
  \global\let\@thanks\@empty
  \global\let\@author\@empty
  \global\let\@date\@empty
  \global\let\@title\@empty
  \global\let\title\relax
  \global\let\author\relax
  \global\let\date\relax
  \global\let\and\relax
  \def\version{\let\version\@version\@gobble}
}
\def\@makepapertitle{%
  \newpage
   \ifnum\draftcontrol=1 {}
   \version\versionno
   \vskip 3em%
   \else
   \hfill\hbox to 3cm {\parbox{4cm}{\@pubnum}\hss}%
   \vskip 3em%
   \fi
   \begin{center}%
   \let \footnote \thanks
     {\LARGE {\@title}}%
     \vskip 1.5em%
     {\normalsize
       \lineskip .5em%
       \begin{tabular}[t]{c}%
         \@author
       \end{tabular}\par}%
     \vskip 1.5em%
     {\@bstract}%
     \end{center}%
     \vskip 1.5em
     \@date%
   \par
}

\gdef\@pubnum{}
\def\pubnum#1{%
  \gdef\@pubnum{#1}}

\gdef\@bstract{}
\def\Abstract#1{%
  \gdef\@bstract{%
   \parbox{\textwidth-0pc}{%
   \centerline{\bf Abstract}\penalty1000%
\kern.2cm%
\noindent
\renewcommand\baselinestretch{1.0}%
{#1}}}
}

\def\ps@paper{\let\@mkboth\@gobbletwo%
     \ifnum\draftcontrol=1
        \def\@oddfoot{\hbox to \textwidth{\tiny \versionno \hfil\tiny\draftdate}%
        \hskip -\textwidth \hbox to \textwidth{\hfil\rm\thepage\hfil}}%
     \else\def\@oddfoot{\hbox to \textwidth{\hfil\rm\thepage\hfil}}
     \fi
     \let\@evenfoot\@oddfoot
}

\def\body{\clearpage
          \pagestyle{paper}
        }

\def\@version#1{\ifnum\draftcontrol=1
\typeout{}\typeout{#1}\typeout{}
\vskip3mm\centerline{\hbox{\fbox{\normalsize{\tt DRAFT -- #1 -- }
                   {\draftdate}}}}\vskip3mm
\fi}
\let\version\@version
\long\def\eqlabel#1{\ifnum\draftcontrol=1
                    \tag@false  
                    \tag*{(\theequation) \hbox to -0.2cm{\hspace{0cm}\small{#1}\hss}}
                    \refstepcounter{equation}
                    \edef\@currentlabel{\theequation}
                    \ltx@label{#1}          
                    \else
                    \label{#1}
                    \fi
                    }
\let\st@bibitem\@bibitem
\let\st@lbibitem\@lbibitem
\ifnum\draftcontrol=1
  \def\@bibitem#1{%
    \st@bibitem{#1}\a@@label{#1}\ignorespaces}
  \def\@lbibitem[#1]#2{%
    \st@lbibitem[#1]{#2}\a@@label{#2}\ignorespaces}
  \def\a@@label#1{%
    \gdef\a@lab{\smash{\normalfont\small#1}}
    \ifvmode
      \if@inlabel
        \global\setbox\@labels\hbox{%
          \llap{\a@lab\let\a@lab\relax
                \kern\@totalleftmargin\kern\marginparsep}%
          \box\@labels}%
      \fi
    \fi}
\fi

\documentclass[12pt,letterpaper]{article}

\usepackage{amsmath,amssymb,array,calc,rotating,epsfig,psfrag}
\usepackage[nosort]{cite}

\ifnum\draftcontrol=1
\fi

\renewcommand\baselinestretch{1.25}
\renewcommand\section{\@startsection {section}{1}{\z@}%
                                   {-3.5ex \@plus -1ex \@minus -.2ex}%
                                   {2.3ex \@plus.2ex}%
                                   {\normalfont\large\bfseries}}
\renewcommand\subsection{\@startsection{subsection}{2}{\z@}%
                                   {-3.25ex\@plus -1ex \@minus -.2ex}%
                                   {1.5ex \@plus .2ex}%
                                   {\normalfont\normalsize\bfseries}}
\renewcommand\subsubsection{\@startsection{subsubsection}{3}{\z@}%
                                   {-3.25ex\@plus -1ex \@minus -.2ex}%
                                   {1.5ex \@plus .2ex}%
                                   {\normalfont\normalsize\it}}
\renewcommand\paragraph{\@startsection{paragraph}{4}{\z@}%
                                   {-3.25ex\@plus -1ex \@minus -.2ex}%
                                   {1.5ex \@plus .2ex}%
                                   {\normalfont\normalsize\bf}}

\numberwithin{equation}{section}

\def\ie{{\it i.e.}}

\def\revise#1       {\raisebox{-0em}{\rule{3pt}{1em}}%
                     \marginpar{\raisebox{.5em}{\vrule width3pt\
                     \vrule width0pt height 0pt depth0.5em
                     \hbox to 0cm{\hspace{0cm}{%
                     \parbox[t]{4em}{\raggedright\footnotesize{#1}}}\hss}}}}

\def\calo         {{\cal O}}

\def\del          {\partial}

\def\sqr#1#2{{\vcenter{\vbox{\hrule height.#2pt
 \hbox{\vrule width.#2pt height#1pt \kern#1pt
 \vrule width.#2pt}\hrule height.#2pt}}}}

\newcommand{\ft}[2]{{\textstyle{\frac{#1}{#2}}}}

\def\bx{\bar{x}}
\def\by{\bar{y}}
\def\bz{\bar{z}}

\def\SU{{\rm SU}}
\def\U{{\rm U}}
\def\dd{\delta}
\def\r{\rho}

\catcode`\@=12

\begin{document}


\title{Radiative Corrections to the Inflaton Potential as an
  Explanation of Suppressed Large Scale Power in Density Perturbations
  and the Cosmic Microwave Background}

\pubnum{%
MCTP-04-57\\
hep-ph/0410117}
\date{October 2004}

\author{
  A.~Buchel$^{(1,2)}$, F.~A.~Chishtie$^{(2)}$, V.~Elias$^{(2)}$,  Katherine Freese$^{(3)}$, R.~B.~Mann$^{(1,4)}$,\\
  D.~G.~C.~McKeon$^{(2)}$, and T.~G.~Steele$^{(5)}$\\[0.4cm]
  \it $^1$Perimeter Institute for Theoretical Physics\\
  \it Waterloo, Ontario N2J 2W9, Canada\\[0.2cm]
  \it $^2$Department of Applied Mathematics\\
  \it University of Western Ontario\\
  \it London, Ontario N6A 5B7, Canada\\
  \it $^3$Michigan Center for Theoretical Physics\\
  \it University of Michigan\\
  \it Ann Arbor, MI 48109, USA\\
  \it $^4$Department of Physics\\
  \it University of Waterloo\\
  \it Waterloo, Ontario N2L 3G1, Canada\\
  \it $^5$Department of Physics $\&$ Engineering Physics\\
  \it University of Saskatchewan\\
  \it Saskatoon, Saskatchewan S7N 5E2, Canada }

\Abstract{ 
The Wilkinson Microwave Anisotropy Probe microwave background data
suggest that the primordial spectrum of scalar curvature fluctuations
is suppressed at small wavenumbers. We propose a UV/IR mixing effect
in small-field inflationary models that can explain the observable
deviation in WMAP data from the concordance model.  Specifically, in
inflationary models where the inflaton couples to an asymptotically
free gauge theory, the radiative corrections to the effective inflaton
potential can be anomalously large. This occurs for small values of
the inflaton field which are of the order of the gauge theory strong
coupling scale. Radiative corrections cause the inflaton potential to
blow up at small values of the inflaton field.  As a result, these
corrections can violate the slow-roll condition at the initial stage
of the inflation and suppress the production of scalar density
perturbations.
}


\makepapertitle

\body

\version\versionno

\section{Introduction}
High precision observational cosmology places tight constraints on the
cosmological parameters of our universe. The Wilkinson Microwave
Anisotropy Probe (WMAP) results measuring anisotropies in the Cosmic
Background Radiation (CBR) \cite{cmb1,cmb2}, when combined with data
on high redshift supernovae and large scale structure, strongly
support the ``concordance model'' for our universe\footnote{The
  concordance model is a spatially flat Universe with an adiabatic,
  nearly scale invariant spectrum of initial fluctuations. In what
  follows we assume a $\Lambda$CDM model as a realization of the
  concordance model.}.  One of the intriguing observations of the CBR
anisotropy spectrum is the suppression of its low-$\ell$ multipoles
compared to the predictions of $\Lambda$CDM model. The low-$\ell$
multipoles of the temperature-temperature (TT) angular power spectrum
correspond to large angular scales --- they encode the information
about the small wavenumbers in the spectrum of primordial density
perturbations. Consequently they provide a window on the detailed
features of the inflaton potential during the part of inflation
\cite{guth} that gives rise to observables in structure formation and
the microwave background, produced roughly 60-50 e-foldings before the
end of inflation.
Unfortunately, it is not possible to disentangle the suppression of
low-$\ell$ modes from cosmic variance limitations. In this paper we
assume the suppression is a physical effect.

Not surprisingly, this suppression of low-$l$ multipoles
(along with the possibility of a
running spectral index) was the focus of much recent
discussion\cite{e1,e2,e3,e4,e5,e6,e7,e8,e9,k,e10,e11,e12,e13,e14,e15,e16}. 
Most previous work
attempted to identify a new ultraviolet (UV) physics responsible for
the suppression of the primordial power spectrum at small wavenumbers.
In this paper we point out a simple low-energy (albeit strongly
coupled) field-theoretic phenomenon that produces a 'feature' in the
inflaton potential that serves to explain the suppression
of low-$\ell$ CBR anisotropies\footnote{Effects of strongly coupled gauge 
theory dynamics on inflation were also studied in \cite{w}.}. 
The observed effect is specific to
small-field inflationary models \cite{r0,r}.  Here, small-field models
are defined to be those models of inflation in which the initial value
of the scalar field is small (e.g. near zero) as it starts rolling
down the potential.

We find a phenomenon that modifies the usual
power spectrum of density fluctuations from inflation,
\begin{equation}
\left(\delta^{(0)}_H\right)^2\propto \left(\frac{k}{k_s}\right)^{n_s-1}\,,
\end{equation}  
to
\begin{equation}
\delta_H^2=\left(\delta_H^{(0)}\right)^2\ f\left(\frac{k}{k_s}\right)\,,
\end{equation}
where the superscript $0$ refers to the unmodified spectrum,
and the form-factor $f(\phi)$ has the following behavior:
\begin{equation}
f(k) \rightarrow 0 \,\,\,\, {\rm (small} \,\, k)
\end{equation}
\begin{equation}
f(k) \rightarrow 1 \,\,\,\, {\rm (large} \,\, k) .
\end{equation}
Here $k_s$ corresponds to the scale of perturbations that are produced
$60$ e-foldings before the end of inflation, which corresponds to the
present horizon size, $k_s\sim (4000 Mpc)^{-1}$.  

The suppression of large-scale power arises because the potential
blows up (becomes infinite) at small values of the field, so that the
field is not slowly rolling at all and production of density
fluctuations is suppressed; see Fig. (1).  This blow-up of the
potential is sharply localized at 60 e-foldings before the end of
inflation, so that ordinary density fluctuations and structure
formation ensue just afterwards (at 60-50 e-foldings before the end).
While the sharpness of the feature is generic to our mechanism, its
location at exactly 60 e-folds before the end requires fine tuning\footnote{
The same fine-tuning problem is acknowledged in all previous work on the small-$\ell$ suppression of CBR
power spectrum we are aware of \cite{e1,e2,e3,e4,e5,e6,e7,e8,e9,k,e10,e11,e12,e13,e14,e15,e16}.
Unfortunately, neither
of the proposed models provided the solution to it.
It would be very interesting 
to identify physical phenomena that alleviates the latter fine tuning problem.}. The
origin of the blow-up of the potential is due to the radiative
corrections for an inflaton coupled to an asymptotically free gauge
theory (analogous to QCD).

\begin{figure}[ht]
\begin{center}
\epsfig{file=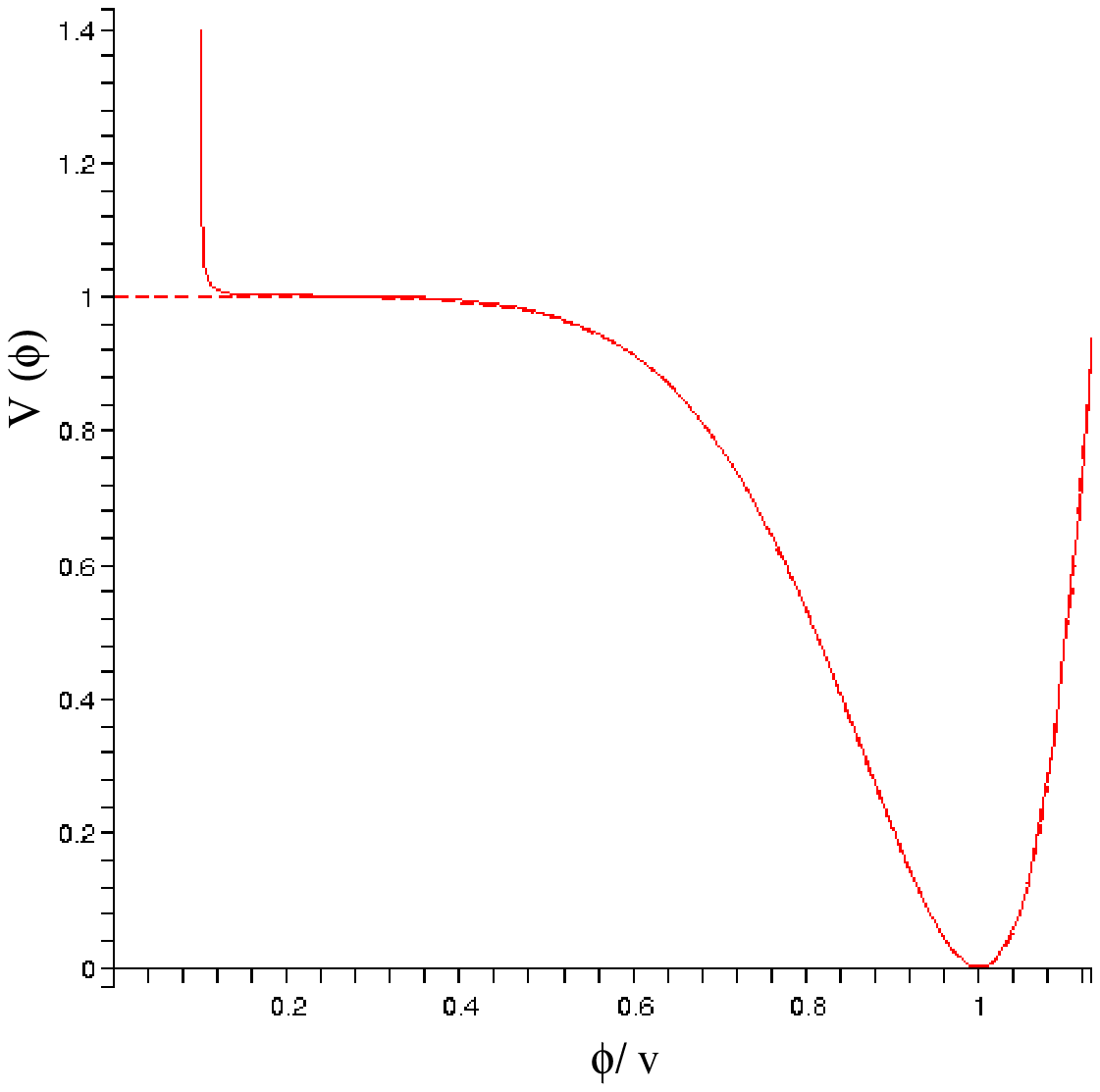,width=0.9\textwidth}
\caption{
  Effective inflaton potential (solid line) in small-field
  inflationary models with inflaton coupled to asymptotically free
  gauge theory. The slow-roll tree-level potential (dashed line)
  receives large radiative corrections whenever the inflaton
  expectation value is close to the gauge theory strong coupling scale
  $\phi\sim \Lambda\ll v$. In the sharply localized regime
  $|\phi-\Lambda|\ll v$ the field is not slowly rolling, and
  production of density fluctuations is strongly suppressed.  }
\label{cases0}
\end{center}
\end{figure}

Thirty years ago Coleman and Weinberg  computed the one-loop effective potential for 
(classically) conformally invariant 
$\SU(2) \times \U(1)$ gauge theory with no quadratic mass term.  However,
due to the massive top quark, the contributions due to the $t$-quark's
Yukawa coupling constant have subsequently been found to dominate over
those of the $\SU(2)\times \U(1)$ gauge coupling constants. Moreover,
when there is a large Yukawa coupling, subsequent leading logarithm
terms to the one-loop effective potential are too large to neglect. In the
next section we review a technique developed in Ref.  \cite{a42,a4}
for all-order summation of leading logarithm terms for the effective
potential, including coupling to asymptotically free theory (QCD).
Radiative symmetry breaking as an explanation for the Higgs mass is
revived in this context, since a prediction for the Higgs mass above
200 GeV can now be accommodated \cite{a42}.

The physics characterizing this model extends to more
general  scalar potentials, including that of any 
inflaton coupled
to an asymptotically free gauge theory.
We explicitly compute the leading contribution to the effective
inflaton (Higgs) potential and show that radiative corrections become
anomalously large for small vacuum expectation values of the inflaton.
This is a reflection of a new UV/IR mixing effect in the model: even
though the inflationary potential has a typical GUT scale height, the
effective QCD coupling in radiative corrections to the inflaton
potential is evaluated at the field value of the inflaton.
The effect is thus most profound if the inflaton value at
the beginning of inflation is close to the gauge theory's strong
coupling scale.  Thus, for a sufficiently small initial inflaton
value, the QCD coupling develops a perturbative Landau pole that
strongly enhances radiative corrections.  

In section 3 we model these radiative corrections in small-field
inflationary scenarios and demonstrate that they suppress the
primordial power spectrum of scalar density fluctuations for small
wavenumbers.  We use the CMBEASY package \cite{cmbeasy} to relate the
latter suppression to the small multipole suppression in the CBR
spectrum of anisotropies as observed by WMAP.  We summarize  our results
in section 4.

\section{QCD contributions to radiative Higgs potential}

A technique for all-order summation of leading logarithm terms for the
effective potential in radiative electroweak symmetry breaking,
including coupling to asymptotically free theory (QCD), has been
developed in Ref.~\cite{a42,a4}. In this section we review this
analysis.

Though our discussion is in the context of radiative symmetry breaking
of the Higgs potential in the Standard Model, the physics of the
ultraviolet-infrared mixing characterizing this model extends to more
general scalar (inflaton) potentials: all that is required is the
coupling of the inflaton to an asymptotically free gauge theory. We
emphasize the latter genericity, as from the inflationary perspective
the radiative symmetry breaking potential discussed in this section is
not suitable for inflation: it predicts unacceptably large amplitude
of density fluctuations.  In the case of the effective radiative
symmetry breaking Higgs potential in the Standard Model the dominant
QCD contribution is given by \eqref{potsin}, which is the main result
of the section. We expect perturbatively divergent contributions of
precisely this kind to be present in all small-field inflationary
models with inflaton coupling  to a QCD-like gauge theory.  Hence, in the
next section we consider an exponential potential for inflation
(different from what is considered in this section), but with the same
physical behavior due to coupling to an asymptotically free theory.

In the absence of an explicit scalar-mass term, the one-loop effective
potential for $\SU(2)\times \U(1)$ gauge theory was computed in 1973
in the seminal paper by Coleman and Weinberg \cite{cw}
\begin{equation}
V_{eff}^{1-loop}=\frac{\lambda\phi^4}{4}+\phi^4\left[\frac{3\lambda^2}{16\pi^2}+\frac{3(3g_2^4+2g_2^2g'^2+g'^4)}
{1024\pi^2}\right]\left(\log\frac{\phi^2}{\mu^2}-\frac{25}{6}\right)
\eqlabel{cw}
\end{equation}
where the $-25/6$ constant is chosen to ensure that 
\begin{equation}
\frac{d^4V_{1-loop}}{d\phi^4}=\frac{d^4V_{tree}}{d\phi^4}=6\lambda
\end{equation}
and $\{g_2,g'\}$ are the $\SU(2)$ and $\U(1)$ coupling constants
respectively.

The above one-loop computation neglects quark Yukawa couplings.
As there is  a heavy $t$-quark, this  is no longer
justifiable\footnote{When there is a  large Yukawa coupling, it is also
  inconsistent to neglect higher-loop contributions to the effective
  potential.}. In fact \cite{a4}, the  Yukawa
coupling of the $t$-quark makes contributions  to the scalar effective potential
$V_{eff}$ which {\it dominate} over those of the $\SU(2)\times \U(1)$ gauge
coupling constants.  As in \cite{a4}, we neglect the $\SU(2)\times
\U(1)$ gauge couplings, and all quark Yukawa couplings except for that of
the $t$-quark.  Thus, the effective potential takes the form
\begin{equation}
V_{eff}=V_{eff}\left(\lambda(\mu),g_t(\mu),g_3(\mu),\phi^2(\mu),\mu\right)\,,
\eqlabel{veff}
\end{equation}   
where $\lambda$ is the quartic scalar-field self-interaction coupling
constant appearing in the tree-level scalar potential
\begin{equation}
V_{tree}=\frac {\lambda}{4}\phi^4+{\rm const}\equiv \frac {\lambda}{4}\phi^4+V^{(0)}\,,
\eqlabel{tree}
\end{equation}
$g_t$ is the Yukawa coupling of the $t$-quark, $g_3$ is the QCD
coupling, and $\mu$ is the renormalization mass scale. The requirement
that $V_{eff}$ is independent of the renormalization scale $\mu$ gives
rise to the familiar renormalization group equation for the effective
potential
\begin{equation}
\begin{split}
0=&\mu \frac{d}{d\mu} V[\lambda(\mu),g_t(\mu),g_3(\mu),\phi^2(\mu),\mu]\\
=&\left(\mu\frac{\del}{\del\mu}+\beta_\lambda\frac{\del}{\del\lambda}+\beta_t\frac{\del}{\del g_t}+\beta_3\frac{\del}
{\del g_3}-2\gamma\phi^2\frac{\del}{\del\phi^2}\right) V\left(\lambda,g_t,g_3,\phi^2,\mu\right)\,,
\end{split}
\eqlabel{pot}
\end{equation} 
where to one loop order in $\lambda$, $g_t$ and $g_3$ 
\begin{equation}
\begin{split}
&\beta_{\lambda}\equiv \mu\frac{d\lambda}{d\mu}=\frac{48\lambda g_t^2}{64\pi^2}+\frac{12\lambda^2}{8\pi^2}
-\frac{3g_t^4}{8\pi^2}+\calo\left(\lambda^kg_{3,t}^{6-2k}\right)\,,\\
&\beta_t\equiv \mu \frac{dg_t}{d\mu}=\frac{\ft 92g_t^3-8g_tg_3^2}{16\pi^2}+\calo\left(\lambda^kg_{3,t}^{5-2k}\right)\,,\\
&\beta_3\equiv \mu \frac{dg_3}{d\mu}=-\frac{7g_3^3}{16\pi^2}+\calo\left(g_{3,t}^{5}\right)\,,\\
&\gamma\equiv -\frac{\mu}{\phi}\frac{d\phi}{d\mu}=\frac{3g_t^2}{16\pi^2}+\calo\left(\lambda^kg_{3,t}^{4-2k}\right)\,.
\end{split}
\eqlabel{1loop}
\end{equation} 
For small coupling constants $\{x,y,z\}$ defined at a scale
$\mu=2^{-1/4}G_F^{-1/2}\equiv v$
\begin{equation}
\begin{split}
&x\equiv g_t^2(v)/4\pi^2\,,\\
&y\equiv \lambda/4\pi^2\,,\\
&z\equiv g_3^2(v)/4\pi^2\,,
\end{split}
\eqlabel{defxyz}
\end{equation}
the summation-of-leading-logarithms effective Higgs potential can be
written as \cite{a4}
\begin{equation}
V_{eff}^{LL}\equiv \pi^2\phi^2S_{LL}=\pi^2\phi^4\left\{\sum_{n=0}^{\infty}x^n\sum_{k=0}^{\infty}y^k
z^l C_{n,k,l}L^{n+k+l-1}\right\}\,,\ (C_{0,0,0}=0)\,,
\eqlabel{lldef}
\end{equation}
where the series $S_{LL}$ is the sum of all contributions involving a
power of the logarithm 
\begin{equation}
L\equiv\ln(\phi^2/\mu^2) 
\end{equation}
that is only one
degree lower than the aggregate power of the couplings $\{x,y,z\}$. In
this approximation the RGE \eqref{pot} takes the following form
\begin{equation}
\left[-2\frac{\del}{\del L}+\left(\frac 94x^2-4 x z \right)\frac{\del}{\del x}+\left(6y^2+3yx-\frac 32 x^2\right)
\frac {\del}{\del y}-\frac{7}{2}z^2\frac{\del}{\del z}-3x
\right]S_{LL}(x,y,z,L)=0\,.
\eqlabel{slle}
\end{equation} 
Remarkably a closed-form solution to \eqref{slle} can be written down \cite{a4} 
\begin{equation}
V_{eff}^{LL}=\pi^2\by(L/2)\bar{\phi}^4(L/2)=\pi^2\by(L/2)\phi^4\exp\left[-3 \int_0^{L/2}\bx(t)dt\right]\,,
\eqlabel{cf}
\end{equation}
where $\{\bx(t),\by(t),\bz(t)\}$ are characteristic functions defined by the differential equations and 
initial conditions
\begin{equation}
\frac{d\bz}{dt}=-\frac 72 \bz^2\,,\qquad \bz(0)=z\,,
\eqlabel{eq1}
\end{equation}
\begin{equation}
\frac{d\bx}{dt}=\frac 94 \bx^2-4\bx\bz\,,\qquad \bx(0)=x\,,
\eqlabel{eq2}
\end{equation}
\begin{equation}
\frac{d\by}{dt}=6\by^2+3\bx\by-\frac 32 \bx^2\,,\qquad \by(0)=y\,,
\eqlabel{eq3}
\end{equation}
\begin{equation}
\frac{d\bar{\phi}}{dt}=-\frac 34 \bx\bar{\phi}\,,\qquad \bar{\phi}(0)=\phi\,,
\eqlabel{eq4}
\end{equation}
Eq.\eqref{eq1} describes the running of the QCD coupling 
\begin{equation}
\bz(t)=\frac{2z}{2+7z\ t}\,.
\eqlabel{zrun}
\end{equation}
For  
\begin{equation}
t\equiv t_s=-\frac {2}{7z}\,,
\eqlabel{ts}  
\end{equation}
the QCD coupling blows-up --- this is the standard IR Landau pole of
the asymptotically free gauge theories. It is straightforward to
evaluate $V_{eff}^{LL}$ in the vicinity of the pole, \ie,
$|L/2-t_s|\to 0$. Given that
\begin{equation}
\begin{split}
&\bx(t)=\frac{4z}{9(2+7z t)}\left[1+\cdots\right]\,,\\
&\by(t)=\frac{\xi z}{2+7z t}\left[1+\cdots\right]\,,\qquad \xi\equiv\frac{\sqrt{689}-25}{36}\,,
\end{split}
\eqlabel{psol}
\end{equation}
where $\cdots$  indicates terms vanishing in the limit $(t-t_s)\to 0$, we find 
\begin{equation}
V_{eff}^{LL}= \frac{\mu^4\pi^2\xi z}{2}\ e^{-\ft {8}{7z}}\ \left(\frac{9x}{2z}\right)^{4/3}\ 
\left(1+\frac 74z\ L\right)^{-25/21}\left[1+  \calo\left(1+\frac 74 z  L\right)\right]
\eqlabel{potsin}
\end{equation}
as $(L/2-t_s)\to 0$, i.e., in the vicinity of the pole.
Notice that the dominant contribution near the perturbative QCD pole
in \eqref{psol}  is insensitive  to the small UV values
of the top quark Yukawa coupling ($x$) and Higgs self-coupling ($y$),
\eqref{eq1}-\eqref{eq3}. On the other hand, the residue of the branch-cut 
in the effective potential \eqref{potsin} does depend on the UV top quark Yukawa 
coupling\footnote{Such dependence arises from the subdominant terms in $\bx$, \eqref{psol}.} as this 
singularity arises {\it only} when this coupling is non-vanishing. 
Leading singular behavior of \eqref{psol}, \eqref{potsin} reflects the fact that an exact
solution of \eqref{eq1}-\eqref{eq3}
\begin{equation}
\begin{split}
\bz(t)=&\frac{2z}{2+7z\ t}\,,\\
\bx(t)=&\frac{4z}{9(2+7z t)}\,,\\
\by(t)=&\frac{\xi z}{2+7z t}\,,
\end{split}
\eqlabel{att}
\end{equation} 
with initial conditions
\begin{equation}
\begin{split}
\bz(0)=z,\qquad \bx(0)=\frac{2z}{9}\,,\qquad \by(0)=\frac{\xi z}{2}\,,
\end{split}
\eqlabel{initatt}
\end{equation} 
is an attractor of the physical\footnote{Subject  to the constraints
  $\{\bx(t),\by(t),\bz(t)\}>0$.}  renormalization group flow. Finally,
the radiative contribution dominance over the tree level Higgs
potential in the full effective potential \eqref{potsin} is sharply
localized near the QCD strong coupling scale.

The {\it existence} of this perturbative singularity is generic for
all asymptotically free gauge theories coupled to an inflaton (Higgs)
field (n.b. the order of the branch-cut and its residue in the
leading-logarithms effective potential \eqref{potsin} is specific to
the Standard Model matter content). The point is simply that radiative
corrections to the tree-level classical inflaton potential become
anomalously large when evaluated for small values of the inflaton
field, a consequence of the perturbative IR pole of the asymptotically
free gauge theory.  In the framework of the effective quantum field
theory, the above conclusion is independent of the scale of the tree-level
inflaton potential, the constant term in \eqref{tree}. In fact, both
$V^{(0)}$ and $v$ (see \eqref{defxyz}) can be of order the GUT scale,
and $V_{eff}$ would still have a perturbative singularity for
sufficiently small $\phi$. Hence in the next section we will turn
to inflation.

\section{UV/IR mixing in small-field inflationary models}

In the previous section we argued that radiative corrections to the
tree-level effective inflaton potential are important, provided the
value of the inflaton is of the same order as the strong coupling scale
of the asymptotically free gauge theory to which it couples.  In what
follows we study a simple model of small-field inflation that
illustrates observed effects on the spectrum of CBR anisotropies.
Here, small-field models are defined to be those models of inflation
in which the initial value of the scalar field is small (e.g. near
zero) as it starts rolling down the potential.  Examples of
small-field models include models based on spontaneous symmetry
breaking phase transitions where the field rolls away from an unstable
equilibrium such as natural inflation \cite{ffo}. In this paper, as a
simple example, we will consider an exponential potential.

We work with the tree level inflaton potential
\cite{i1,i2,i3}
\begin{equation}
V_{tree}=V_0 e^{-\lambda \phi}\,,
\eqlabel{ipot}
\end{equation}
where $\phi\equiv \Phi/m_{pl}$ is the inflaton field in Planck units.
We assume that the inflaton couples to some  asymptotically free gauge
GUT.  As a result, the tree-level inflaton potential will receive
radiative corrections.  We model radiative corrections to \eqref{ipot}
as
\begin{equation}
V_{eff}=V_{tree}+V_{radiative}\equiv V_0 e^{-\lambda \phi}+\alpha\ \frac{V_0}{\ln \frac{\phi}{\Lambda}}\,, 
\eqlabel{vefff}
\end{equation}   
where $\alpha$ is proportional to the coupling constant of the GUT
asymptotically free gauge theory at the GUT scale, thus $0<\alpha\ll
1$. The quantity $\Lambda$ is the strong coupling scale of the gauge
theory in Planck units, $\Lambda\ll 1$. 
The form of the potential is similar to that of Fig. (1);
the potential becomes infinite at $\phi \rightarrow \Lambda$.
The starting point for inflation then clearly has to
be at $\phi=\phi_i>\Lambda$. 

The effective potential $V_{eff}$ has the required features to explain
the suppression of the observed CBR anisotropy spectrum at small $l$.
Indeed, if the beginning of inflation\footnote{By the 'beginning' we
  mean the very outset  of the $\sim 60$ e-foldings characterizing  inflation.}  $\phi_i$ is close to $\Lambda$, \ie,
$(\phi_i-\Lambda)\ll \Lambda$, the radiative contribution in
\eqref{vefff} is dominant. The inflaton then rolls too fast to
effectively generate primordial density perturbations at small
wavenumbers.  This will further translate into the suppression of the
CBR anisotropy spectrum at large angles (or small $l$). This
suppression is rather sharply localized: for $\phi \sim 2 \Lambda$ or
larger, the radiative contribution does not exceed $V_{tree}$:
\begin{equation}
V_{radiative}\propto \alpha\ V_0\ll V_0\approx V_{tree}\,.
\eqlabel{vrad}
\end{equation}

We now turn to the quantitative analysis of inflation with \eqref{vefff}. First, to determine the scales and parameters 
we set $\alpha=0$, thus $V_{eff}=V_{tree}$. The standard slow roll parameters are\footnote{Prime denotes 
derivative with respect to $\Phi$.}
\begin{equation}
\begin{split}
&\epsilon=\frac{m_{pl}^2}{2} \left(\frac{V_{eff}'}{V_{eff}}\right)^2=\frac 12 \lambda^2\,,\\
&\eta=m_{pl}^2\left(\frac{V_{eff}''}{V_{eff}}\right)=\lambda^2\,,
\end{split}
\eqlabel{sr}
\end{equation}  
producing scalar density fluctuations with power spectrum 
\begin{equation}
\delta_H^2\bigg|_{\alpha=0}\equiv \left(\delta^{(0)}_H\right)^2\propto \left(\frac{k}{k_s}\right)^{n_s-1}\,,
\eqlabel{ps}
\end{equation}  
with spectral index 
\begin{equation}
n_s-1=-6\epsilon+2\eta=-\lambda^2\,.
\eqlabel{nsdef}
\end{equation}  
In \eqref{ps} $k_s$ corresponds to the scale of perturbations that
are produced $60$ e-foldings before the end of inflation, which
corresponds to the present horizon size, $k_s\sim (4000 Mpc)^{-1}$.
Within the slow-roll approximation, adiabatic density perturbations
are given by
\begin{equation}
\delta_H^{(0)}\sim \frac{1}{\sqrt{75}\pi m_{pl}^3} 
\frac{V_{eff}^{3/2}}{V_{eff}'}=\frac{V_0^{1/2}}
{\sqrt{75}\pi m_{pl}^2\lambda}\ e^{-\lambda\phi/2}\,.
\eqlabel{dpert}
\end{equation}
This quantity should equal $1.91\cdot 10^{-5}$ at about $N_e\sim 60$ e-foldings before the end of inflation. 
Assuming $\lambda\phi_i\ll 1$ and $m_{pl}=2.4\cdot 10^{18}$ GeV, 
for $\lambda\sim 0.1$ we find that $V_0^{1/4}\sim 10^{16}$ GeV from \eqref{dpert}. If inflation ends at $\phi=\phi_f$, the number of e-foldings $N_e$ is  
\begin{equation}
N_e=-\frac{1}{m_{pl}^2}\int_i^f\ \frac{V_{eff}}{V_{eff}'}\ d\Phi=\frac 1\lambda\int_i^f d\phi=\frac{\triangle\phi}{\lambda}\,.
\eqlabel{ndef}
\end{equation}
Thus to get sufficient inflation, $\triangle \phi\sim 10$. Since
derivatives with respect to $\phi$ can be expressed with respect to
wavenumber $k$ as 
\begin{equation}
\frac{m_{pl}}{2}\sqrt{\epsilon}\frac{d}{d\Phi}=(1-\epsilon) \frac{d}{d\ln k}\,,
\eqlabel{derex}
\end{equation} 
or 
\begin{equation}
\frac{\lambda}{2^{3/2}} \frac{d}{d\phi}\approx  \frac{d}{d\ln k}\,,
\eqlabel{derex1}
\end{equation} 
we can relate the expectation value of the inflaton and the scale of
perturbations $k$ leaving the horizon at the corresponding instant
during inflation
\begin{equation}
\phi(k)\approx\phi_i+\frac{\lambda}{2^{3/2}} \ln \frac{k}{k_s}\,.
\eqlabel{derex2}
\end{equation}

Upon inclusion of the radiative correction to the tree-level potential
\eqref{ipot},  the primordial spectrum of density perturbations
\eqref{ps} will be modified according to
\begin{equation}
\delta_H^2=\left(\delta_H^{(0)}\right)^2\ f\left(\frac{k}{k_s}\right)\,,
\eqlabel{sm}
\end{equation}
where 
\begin{equation}
\left(\delta_H^{(0)}\right)^2\sim \frac{1}{\sqrt{75}\pi m_{pl}^3} 
\frac{V_{tree}^{3}}{V_{tree}'^2}
\eqlabel{old}
\end{equation}
and the form-factor $f(\phi)$ can be estimated using the slow-roll
approximation in Eq.(\ref{dpert}) and Eq.(\ref{vefff}),
\begin{equation}
f\approx \frac{\left(1+\frac{V_{radiative}}{V_{tree}}\right)^3}{\left(1+\frac{V_{radiative}'}{V_{tree} '}\right)^2}
=\frac{\left(1+\frac{1}{\ln\frac{\phi}{\Lambda}}\ {\alpha e^{\lambda\phi}}\right)^3}{
\left(1+\frac{1}{\frac{\phi}{\Lambda}\ln^2\frac{\phi}{\Lambda}}\ \frac{\alpha e^{\lambda\phi}}{\lambda\Lambda}
\right)^2
}\,.
\eqlabel{slow}
\end{equation}
For inflaton expectation values $\phi$ close to the gauge theory 
strong coupling scale $\Lambda$ (where $V_{radiative} >> V_{tree}$), we find that
\begin{equation}
f(\phi)\sim \alpha\lambda^2\Lambda^2\ \ln \frac{\phi}{\Lambda}\,,\qquad (\phi-\Lambda)\ll \Lambda\,.
\eqlabel{small}
\end{equation}
As $\phi/\Lambda\sim \calo(1)$, $f(\phi)$ rapidly approaches zero. More
precisely, assuming $\frac{\alpha}{\lambda\Lambda}>1$, the transition
between the small $\phi$-regime \eqref{small} and $f(\phi)=1+\calo(1)$
occurs within 
\begin{equation}
\triangle\left(\frac{\phi}{\Lambda}\right)\sim \frac{\alpha}
{\lambda\Lambda}\equiv \zeta\,
\eqlabel{tranlength}
\end{equation}
or, equivalently, using \eqref{derex2},
\begin{equation}
\triangle\left(\frac{k}{k_s}\right) \sim
\delta\equiv \exp\left(\frac{2^{3/2}\zeta\Lambda}{\lambda}\right)
=\exp\left(\frac{2^{3/2}\alpha}{\lambda^2}\right).
\end{equation}
With $\phi_i\sim \Lambda$, and using \eqref{derex2}, we can approximate
\begin{equation}
\begin{split}
&f\left(\frac{k}{k_s}\right)=0,\qquad \frac{k}{k_s}<1\,,\\
&f\left(\frac{k}{k_s}\right)=\frac{\ln\frac{k}{k_s}}{\ln\delta},\qquad  1\le \frac{k}{k_s}\le \delta\,,\\
&f\left(\frac{k}{k_s}\right)=1,\qquad \frac{k}{k_s}>\delta\,,
\end{split}
\eqlabel{apf}
\end{equation}
Equivalently, using
Eq.(\ref{derex2}), the form-factor in the nontrivial region of
\eqref{apf} can be approximated as having a linear dependence on
$\phi$,
\begin{equation}
f(\phi)= \frac{\phi-\Lambda}{\Lambda\zeta},\qquad 0\le(\phi-\Lambda)\le \Lambda\zeta\,.  
\end{equation}
Given the modified primordial power spectrum \eqref{sm} (with \eqref{ps},\eqref{nsdef} and the form-factor 
\eqref{apf}), the present day power spectrum $P(k)$ is obtained as 
\begin{equation}
\frac{k^3}{2\pi} P(k)=\left(\frac{k}{a H_0}\right)^4\ T^2(k)\ \delta^2_H(k)\,,
\eqlabel{present}
\end{equation}  
where $T(k)$ is a transfer function as in \cite{i1}, and $H_0$ is the
present Hubble value.  

In Fig.~(2) we show the result generated with
the CMBEASY program \cite{cmbeasy}, which provides the power spectra
conversion \eqref{present}. In the latter analysis we take
$\lambda=.2$ and adjust cosmological parameters as given by the best
fit model of the WMAP collaboration. The two solid lines correspond to
a choice of $\delta=\{2,4\}$ in \eqref{apf}. The dashed line
corresponds to turning off the radiative correction, \ie, setting
$\alpha=0$.  For $\delta=4$ we find from \eqref{apf} $\alpha\approx
0.02$. Thus consistency of above approximations, \ie, $\xi>> 1$
requires $\Lambda\ll 10^{-1}$. The latter is consistent with the
assumption that the strong coupling scale of the GUT gauge theory is
below the Planck scale.
   
We comment here on the validity of our approximation. 
Small-$k$ suppression form-factor \eqref{apf} has been obtained in the slow-roll approximation, 
which strictly speaking, is violated at the initial stages of inflation. Alternatively, one 
can argue that the initial stage of inflation can be described by ``kinetic regime''\footnote{This is inflationary 
stage where the dominant contribution to the inflaton energy comes from its kinetic energy.}, 
where the suppression form-factor was found (numerically) to be 
\begin{equation}
f\left(\frac{k}{k_s}\right)\propto \left(\frac{k}{k_s}\right)^a
\eqlabel{sli}
\end{equation} 
where $a\approx 3.35$ \cite{e6}. Though result \eqref{sli} was obtained in the context of chaotic (large-field) inflation,
it is straightforward to see that it is universal as long as inflaton energy is predominantly kinetic. 
Indeed, the primordial power spectrum is 
\begin{equation}
\dd_H\simeq \left(\frac{\dd\rho}{\r+p}\right)_{k=a H}
\eqlabel{rev1}
\end{equation}
where in the kinetic regime (assuming $\dd\dot\phi=\ft{d}{dt}\dd\phi$)
\begin{equation}
\dd\r \simeq \dot\phi\ \dd\dot\phi\simeq \dot\phi\ \frac{\dot H}{2\pi}
\eqlabel{rev2}
\end{equation} 
Using $\r+p=-\dot H m_{pl}^2/(4\pi) $, we find 
\begin{equation}
\dd_H\propto -\frac{\dot\phi}{m_{pl}^2}\bigg|_{k=a H}
\eqlabel{rev3}
\end{equation}   
where the cosmic time dependence in $\dot\phi =\frac{d}{dt}\phi(t)$ is assumed to be eliminated through the horizon crossing 
condition for the comoving momentum $k$ as $k=a(t) H$. In the kinetic regime $\dot\phi^2\gg V(\phi)$, thus 
expansion of the Universe is well approximated by a fluid with equation of state $p=\rho$. For such an expansion we have 
$\r\simeq \frac 12 \dot\phi^2 \propto a^{-6}$ and $a\propto t^{1/3}$, which in turn implies 
that $\dot \phi\propto -t^{-1}$. From the Friedmann equation we conclude that $H\propto \rho^{1/2}\propto t^{-1}$, and thus the comoving scale 
$k$ crossing the horizon at cosmic time  $t$ is $k=a(t) H\propto t^{-2/3}$. Comparing the latter scale with the $\dot\phi$
time evolution we conclude 
\begin{equation}
\dot\phi\propto - k^{3/2}
\eqlabel{rev4}
\end{equation}
leading to 
\begin{equation}
\dd_H^2\propto \left(\frac{k}{k_s}\right)^3
\eqlabel{rev5}
\end{equation}
which approximately reproduces numerical analysis of \cite{e6}. Even though the suppression form-factor in the primordial power
spectrum \eqref{rev5} has power-law rather than logarithmic behavior (as in \eqref{apf}), we find that the corresponding CBR spectrum of anisotropies 
is very similar to the one presented in Fig.2 (also compare with \cite{k}).

\begin{figure}[ht]
\begin{center}
\epsfig{file=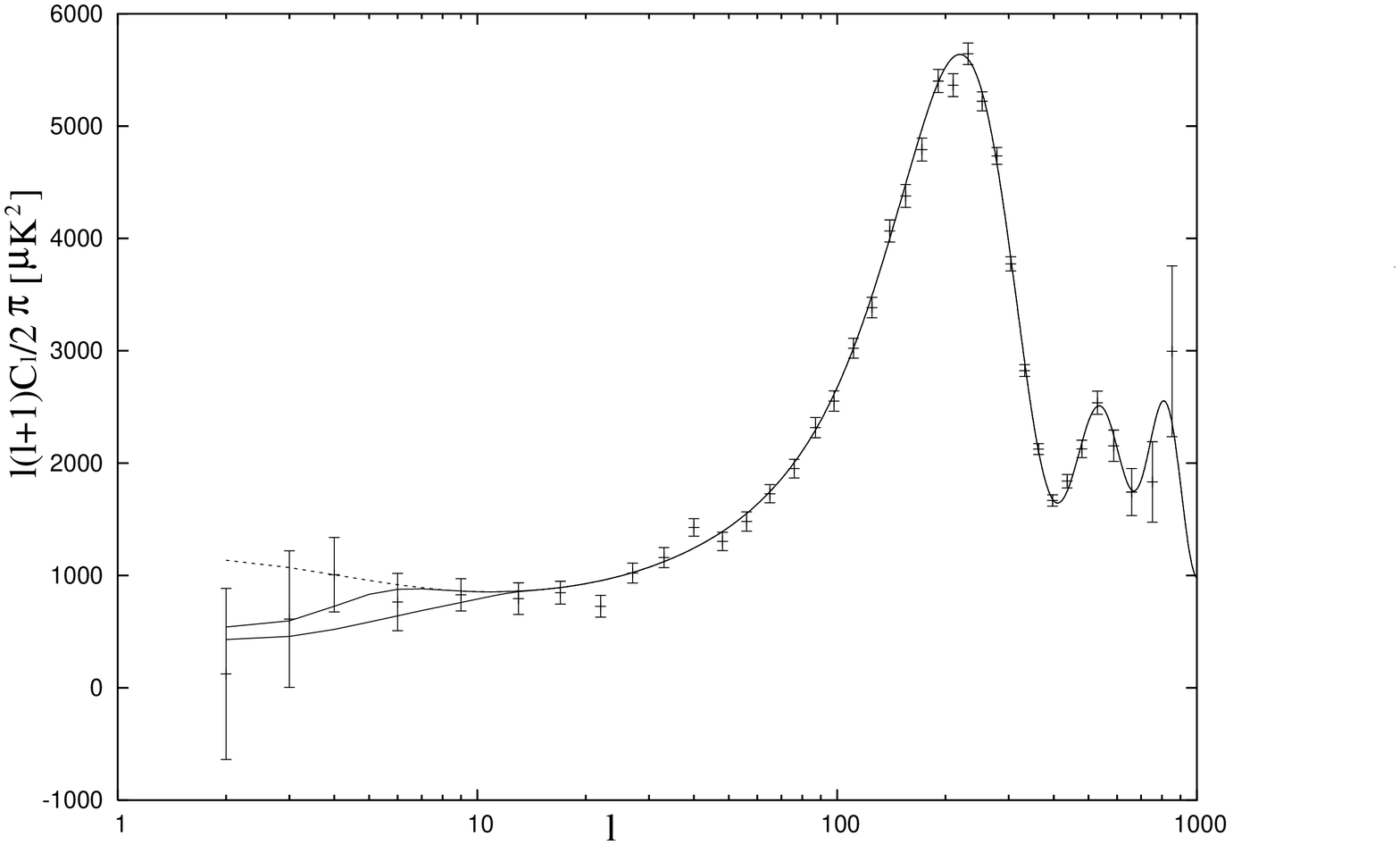,width=1.3\textwidth}
\caption{
  CBR spectrum of anisotropies with the primordial spectrum of scalar
  density fluctuations for the effective inflaton potential
  \eqref{veff}. We set $\lambda=0.2$, $k_s=0.00025\ Mpc^{-1}$.  The
  scale of the tree level potential is $V_0^{1/4}\sim 10^{16}$ GeV.
  Two solid lines correspond to $\delta=\{2,4\}$ in the form-factor
  \eqref{apf}, with larger $\delta$ corresponding to stronger
  suppression at small $\ell$.  The dashed line corresponds to the
  anisotropy spectrum from a tree-level approximation in the effective
  potential \eqref{veff}, \ie, setting $\alpha=0$. Cosmological
  parameters are as given by the best fit of the WMAP collaboration.  }
\label{cases}
\end{center}
\end{figure}

\section{Summary}
WMAP data indicates the suppression of small-$\ell$ multipoles in the
anisotropy power spectrum of the CBR. It is difficult to isolate this
suppression from the cosmic variance limitations. However, when taken
seriously, such suppression can be due to a suppression of
power in the spectrum of primordial scalar density fluctuations during
inflation.  As the generation of quantum fluctuations during inflation
is most efficient in the slow-roll regime, the CBR anisotropy
measurements provide a signature for the breakdown of the slow-roll
conditions during the first few e-foldings of the single field
inflationary models. In this paper we have identified a low-energy
field-theoretic phenomenon that arises generically in small-field
inflationary models in which the inflaton is coupled to an
asymptotically free GUT-like gauge theory. Here, the small vacuum
expectation values of the inflaton field probe the strongly coupled
dynamics of this gauge theory. This produces large radiative
corrections to the tree-level inflaton potential which break its
slow-roll conditions.  The effect is most profound if the inflaton
scale at the beginning of inflation is close to the gauge theory's
strong coupling scale.

In this paper we have used summation of leading logarithms in the
perturbation theory as a model of a strongly coupled dynamics of an
inflaton and asymptotically free gauge theory.  It would be very
interesting to explore nonperturbative effects of the quantum field
theory in this setting, particularly space-time curvature corrections.

\section*{Acknowledgments}
We would like to thank R.~Brandenberger, R.~Dick, E.~Gorbar, V.~A.~Miransky for
helpful conversations.  We are grateful to C.~M.~Mueller for help with
CMBEASY, and W.~H.~Kinney for comments on a manuscript.  
Research at Perimeter Institute is supported in part by
funds from NSERC of Canada. AB, FAC, VE, RBM, DGCM, and TGS  acknowledge support by an NSERC
Discovery grant. KF thanks the Michigan Center for Theoretical Physics
and the DOE via the University of Michigan for support.
AB and KF would like to thank Aspen Center for Physics for
hospitality where part of this work was done.

\end{document}